Examining bias perpetuation in academic search engines: an algorithm audit of Google and Semantic Scholar

Kacperski, Celina[1,2]\*; Bielig, Mona[2]; Makhortykh, Mykola[3]; Sydorova, Maryna[3]; Ulloa, Roberto[1,4]\*

[1] Konstanz University, Germany

[2] Seeburg Castle University, Austria

[3] Bern University, Switzerland

[4] GESIS – Leibniz Institute for the Social Sciences, Germany

Corresponding author: Roberto Ulloa, roberto.ulloa@gesis.org

\*The first and last author share co-first authorship on this publication.

**Abstract**

Researchers rely on academic web search engines to find scientific sources, but search engine mechanisms may selectively present content that aligns with biases embedded in the queries. This study examines whether confirmation biased queries prompted into Google Scholar and Semantic Scholar will yield results aligned with the query's bias. Six queries (topics across health and technology domains such as 'vaccines', 'internet use') were analyzed for disparities in search results. We confirm that biased queries (targeting 'benefits' or 'risks') affect search results in line with the bias, with technology-related queries displaying more significant disparities. Overall, Semantic Scholar exhibited fewer disparities than Google Scholar. Topics rated as more polarizing did not consistently show more disparate results. Academic search results that perpetuate confirmation bias have strong implications for both researchers and citizens searching for evidence. More research is needed to explore how scientific inquiry and academic search engines interact.

**Introduction**

Belief development involves a series of cognitive steps, including hypotheses formation, information acquisition, and belief adjustments based on the found information (Klayman, 1995). In the field of information retrieval, the impact of cognitive biases on information search behavior has been well-documented (for a review, see Azzopardi, 2021). Confirmation bias is the tendency to actively seek, interpret, and remember information that validates one's preexisting beliefs while disregarding contradictory information (Klayman, 1995; Nickerson, 1998; Oswald & Grosjean, 2004). Confirmation bias has been shown to significantly affect information search behavior (Azzopardi, 2021; Knobloch-Westerwick et al., 2015), particularly when using search engines in an online context, with participants actively querying and selecting information from search results that affirmed their preconceived beliefs (Vedejová & Čavojová, 2022; White, 2013).

Individuals' confirmation bias might be inadvertently perpetuated and amplified by search engine biases: search engines often selectively present certain content to users, and in a skewed manner, i.e., in deviation from a unbiased baseline (Goldman, 2005; Vaughan & Thelwall, 2004; White, 2013). As an example, content bias is a deviation of a search engine index from known evidence distributions: in an audit of health related queries, search engine results skewed towards benefits (White & Hassan, 2014) and audits of gender representation show a skew to the detriment of women (Kay et al., 2015; Ulloa, Richter, et al., 2022). Search engine biases have been a topic of much recent attention, as search engines ranking and filtering mechanisms are considered the most influential information gatekeepers in today's online media environment (Laidlaw, 2010; Wallace, 2018). Algorithmic curation has been shown to affect users' beliefs and decisions, including through feedback loops with existing biases such as the confirmation bias (Noble, 2018; White & Horvitz, 2015). Preliminary experimental evidence with curated content in a custom search engine suggests that search results for queries with explicit inherent directionality of a belief (e.g., positive vs negative framing of coffee's relation with hypertension) will display results confirming the directionality (Kayhan, 2015) – however, this has not been shown with results from major scholarly search engines.

Information search biases are especially detrimental in scientific practice. Due to the overwhelming magnitude of information on the internet, heuristic information search is used by scientists to help formulate research questions and hypotheses – this means using previously known (sometimes biased) information as a shortcut to find supporting evidence (Phillips et al., 2002), and using simple search, where search stops at the first cue that supports a pre-existing notion (Dhami & Harries, 2010). The manner in which academic web search engines output sources and evidence on a variety of scientific topics is therefore highly relevant (Ortega, 2014). Major academic engines, such as Google Scholar or Semantic Scholar, commonly employ algorithms to govern the prioritization and display of academic literature search results for users. This has implications for the interactions with their sources, because the metric used for sorting ('relevance') is not well-explained (Jordan & Tsai, 2022), though there is evidence that citation count and query matching are the most influential factors (Rovira et al., 2018). Previous analyses have suggested that many academic search engines, including Google Scholar, might not be usable in systematic search due to issues with precision, query functionality and reproducibility (Gusenbauer & Haddaway, 2020).

Considering that scholars are human (Veldkamp et al., 2017) and therefore have human biases (Epstein, 2004; Hergovich et al., 2010; Masnick & Zimmerman, 2009), confirmation bias together with resulting algorithmic output sorted by relevance may lead to distortions in how researchers develop scientific theory (Greenwald et al., 1986; Kayhan, 2015; Schumm, 2021), i.e. there exists a risk that algorithms may amplify existing biases by focusing on relevance alone: a first biased search query might be exacerbated if search engines feedback the query back with sources skewed towards the bias (White, 2013; White & Hassan, 2014). The consequences of diminished trust in science (Vranic et al., 2022) make it imperative to critically examine the operational principles of such algorithms and consider how they impact research (Jordan, 2022), and how research in academic search engines can be encouraged so that scientists' biases are reduced and not exaggerated (von Hippel & Buck, 2023). This is where we position our contribution, to answer the question how biased query formulation affects results displayed in academic search engines.

A search for scientific sources is especially critical when it comes to topics of high public interest (e.g., related to health or technology use) and those that highly polarize individuals (e.g., vaccines, cryptocurrency use). Non-scientists who are looking to inform themselves on topics relevant to their views and identities are affected, for example, parents with anti-vaccination beliefs. Led by confirmation bias, such a subsample of participants searched for evidence regarding vaccination only using negative phrases and questions (Wharton-Michael & Wharton-Clark, 2020). In another study, when searching for sources on polarizing topics (e.g., death penalty), participants showed a more pronounced confirmation bias than those searching for the control topics (e.g., school uniforms) (Vedejová & Čavojová, 2022). However, following the recent uptick in scrutiny towards biases in algorithmic search (Noble, 2018; Ulloa, Makhortykh, et al., 2022), at least Google has attempted to correct results disparities to showcase a more consistent results section, especially across queries of high interest or relevance (Google, 2021; Ulloa, Richter, et al., 2022). We explore therefore whether for more polarizing and salient topics such as vaccination, biased queries would lead to different result displays compared to neutral topics.

In summary, it is important to understand not only how academic search engines display results when confronted with queries biased towards confirmation or disaffirmation, but also whether they differ regarding domains and topics, specifically the level of polarization of the topics.

**Hypotheses**

We expect that querying an academic search engine for evidence of an issue's benefits will bias search results towards confirming benefits, and asking for an issue's risks will bias academic results towards confirming risks: we will in the following name the difference between the two "disparity". We are investigating the search results disparity, and how pronounced it is across domains, topics and search engines.

We pose the following hypotheses:

**H1:** Biased queries will increase results disparity.

**H2**. Results disparity will differ across different domains, i.e., health versus technology, and across topics.

**H3:** Results disparity will differ across academic search engines (Google Scholar vs Semantic Scholar).

**H4.** Results disparity will be greater for topics that are rated as more polarizing compared to those rated as more neutral.

## Methods

**Query selection**

We experimentally tested academic search engine results with 12 search queries: risks and benefits of health issues (coffee, vaccines, covid vaccines) and technology use issues (internet, social media, cryptocurrencies), see Table 1. The table presents the 12 queries used to prompt the search engine. The first column indicates the domain in which each query is allocated, the second, the topic of the query and the third and fourth, the query used for prompting risks and benefits respectively. We point out that we did not include a neutral "unbiased" control query, as we found ourselves challenged defining what would be a neutral query in this context. This will be further discussed in the limitations section.

*Table 1. Overview of queries selected for category and topic, with risks and benefits.*

| domain | topic | risks query | benefits query |
|---|---|---|---|
| health | coffee | coffee consumption health risks | coffee consumption health benefits |
| | vaccines | vaccines health risks | vaccines health benefits |
| | covid vaccines | covid vaccines health risks | covid vaccines health benefits |
| technology | internet | internet use risks | internet use benefits |
| | social media | social media use risks | social media use benefits |
| | cryptocurrency | cryptocurrency use risks | cryptocurrency use benefits |

Our queries were selected to showcase, per domain, one non-polarizing topic (coffee, internet) and two topics that were polarizing, which we verified using a survey conducted on Prolific.com with crowd workers (N=72). For this, participants were asked to rate the 6 topics first on how polarizing they were ("How polarizing are the following topics? Polarizing in this case means that there is public and/or political disagreement about it and that evidence for or against it is seen as contentious." 0-100 slider scale, no default) and how salient they were perceived in the media ("How salient are the following topics? Salient in this case means how commonly these topics appear on general media (news, TV, social media)." 0-100 slider scale, no default). Participants' ratings supported our topic selection (see Results section).

**Data collection**

To carry out the data collection, we used 2 search engines: Google Scholar and Semantic Scholar. We chose these search engines due to, for Google, its market dominance (Nicholas et al., 2017) and for Semantic, its LLM-powered semantic analysis layer (Jones, 2015), with the hypothesis that they would differ in results displayed. We had access to 2 regions: Frankfurt (German) and Ohio (USA); and used 2 web browsers: Firefox and Chrome. We employed e2-standard-2 virtual machines with 2 CPUs, 8GB of RAM, Debian Linux OS provided by Google Compute Engine. We employed a total of 16 virtual machines (8 in each region), resulting in a total of 32 browsers used for the experiments.

We followed Ulloa et al. (2022) in our algorithm auditing protocol, but we employed Selenium for headless browser control and data storing (instead of browsing extension). The data collection process spanned 8 days, from September 1st to September 8$^{th}$, 2022. Each day, we initiated the virtual machines at 9 AM CET and captured the HTML corresponding to the search results pages of each experimental condition (query, engine, region and browser) four times. Given the 12 queries, 8 days, 2 engines, 2 regions, 2 browsers and 4 repetitions, we expected 3072 HTML pages. As the process is not technically perfect, as outlined in learnings on algorithm auditing (Ulloa, Makhortykh, et al., 2022), we failed to collect 217 pages (7.13%), 27 due to broken HTML on Semantic Scholar. We obtained at least 2 copies of each condition, except for the 2nd day of data collection (September 2$^{nd}$, 2022) for the query "coffee consumption health risks" in Semantic Scholar (with 0 results). A full table of coverage is presented in the Supplementary Materials S1.

For pre-processing, the HTML of each search page was parsed to extract the top-10 article results of every search query. For each engine, we extracted the unique results using the title, the hyperlink and the authors, resulting in 108 articles for Semantic Scholar and 139 for Google Scholar. We scraped the hyperlinks to retrieve the page corresponding to each of the unique articles and parsed the HTML to obtain the potential abstract of each article (104 for Google Scholar, and 69 for Semantic Scholar). We manually checked, corrected, and completed all 247 abstracts and removed duplicates. 242 abstracts were annotated. One of the authors manually annotated all abstracts, while the other authors split 60 equally among themselves to check inter-annotator reliability, which was very high (Cohen's Kappa = .8; Krippendorff's Alpha = .85; Brennan-Prediger = .8). Abstracts where discrepancies occurred where checked and corrected with consensus if necessary. Annotation protocol was to rate whether the article confirmed risk and/or benefits of each of the topics (see Table 1). The articles were then assigned a valence of 1 if the article only confirmed benefits (n=57), -1 if the article only confirmed risks (n=35), 0 if the article confirmed both risk and benefits (n=74), and 0 if the article confirmed neither risks nor benefits (n=69). 6 articles were tagged with missing values as they were rated as not related to the topic at all.

**Data analysis**

Our final dataset comprised as response variable the rated valence per article, and our predictors – the query bias (risk, benefit), the domain (health, technology), individual queries (coffee, vaccines, covid vaccines, and internet, social media, cryptocurrencies), search engine (Google Scholar, Semantic Scholar), location (US, Germany) and browser (Firefox, Chrome). We did not find significant differences for locations and browsers in our analyses and will therefore not further report on them. The data also contained meta information, including an abstract ID and the date of data collection. We used the lm package in R (R Development Core Team, 2008) to conduct a linear regression predicting the abstract valence from query bias, domain/topic and search engine, as well as all interactions between the three, controlling for region and browser as well as random intercepts for individual queries, abstract ID and date. Emmeans (Lenth, 2023) was used for contrasts. We report standardized betas as effect size estimates and p-values for significance testing. Full model result outputs can be found in the supplementary materials.

## Results

**Model results**

We hypothesized that queries with a bias towards benefits and risks of issues would elicit search result disparities matching the query bias. Figure 1 displays abstract valences as a result of query bias (risk/benefit, red/blue), domain (health/technology, top/bottom) and search engine (Google/Semantic Scholar, left/right) for the six queries (coffee, vaccines, covid vaccines, internet, social media, cryptocurrencies).

FIGURE 1

Table 2 showcases statistical effects and significances of the reported relationships. We confirm H1, and find that query bias positively predicts abstract valence, so that queries with benefits elicit more results reporting benefits and queries with risks elicit more results reporting risks, i.e., overall results disparities are statistically significant. In line with H2, we also find a statistically significant interaction of domain with query bias, with technology related queries more likely to show disparities. Finally, in line with H3, we find an interaction of search engine with query bias: Semantic Scholar results are overall less likely to show disparities than Google Scholar results.

*Table 2. Resulting outputs from the linear regression model. We report effect of query (treatment: benefits vs risks), in interaction with the domain (technology vs health) and search engine (Google vs Semantic).*

| Hypotheses | Effect on valence rating (disparity) | Estimate (b) | p-value | Confidence Intervals 95% |
|---|---|---|---|---|
| H1 | query | 0.195 | p <.001 | [0.168, 0.219] |
| H2 | query * domain | 0.322 | p <.001 | [0.288, 0.359] |
| H3 | query * search engine | -0.254 | p <.001 | [-0.249, -0.178] |
|  | query * domain * search engine | 0.479 | p <.001 | [0.430, 0.529] |
| **Contrasts on specific subtopics** | | | | |
| H4 | Google – coffee | -0.170 | p <.001 | [-0.222, -0.117] |
|  | Google – vaccines | 0.600 | p <.001 | [0.547, 0.653] |
|  | Google – covid vaccines | 0.148 | p <.001 | [0.093, 0.203] |
|  | Semantic – coffee | 0.111 | p <.001 | [0.050, 0.171] |
|  | Semantic – vaccines | 0.07 | p <.001 | [0.013, 0.127] |
|  | Semantic – covid vaccines | -0.333 | p <.001 | [-0.389, -0.277] |
|  | Google – internet | 0.546 | p <.001 | [0.493, 0.599] |
|  | Google – social media | 0.300 | p <.001 | [0.247, 0.353] |
|  | Google – crypto | 0.709 | p <.001 | [0.656, 0.762] |
|  | Semantic – internet | 0.600 | p <.001 | [0.545, 0.655] |
|  | Semantic – social media | 0.770 | p =.006 | [0.714, 0.827] |
|  | Semantic – crypto | 0.867 | p <.001 | [0.811, 0.924] |

The triple interaction is positive and significant, which requires a more thorough investigation of the individual queries, and patterns of relationships between our variables, as seen in Figure 1. We thus further investigate results for the individual health and technology topics in the respective individual search engines and domains.

First, for the health domain and Google Scholar, we find only small results disparity for both the coffee query and for covid vaccines query. We do find a larger disparity for vaccines in general. For Semantic Scholar, instead, the largest results disparity exists for covid vaccines, however, here, searching for risks displays more benefits than searching for benefits. The results disparities for coffee and vaccines are very small.

Secondly, for the technology domain, there is a strong tendency to display more risks when risks are being searched for, and results disparities are large across all queries. For Semantic Scholar, the size is consistent across all three queries, though largest for cryptocurrencies. For Google Scholar, the results disparity is also large for internet and cryptocurrency, but smaller for social media.

To summarize, we don't find a clear pattern that would provide evidence to support H4, i.e., we don't see a consistent increase in results disparities for polarizing topics such as vaccines and covid vaccines, though we don't find reduced results disparities either.

**Verification of polarization of topics**

Results of the survey to rate polarization and salience of our selected topics are displayed in Figure 2 and confirm our selection for coffee and internet as neutral topics. The topics vaccines (vaccines: $b= 40.97$, $p <.001$; covid: $b= 48.92$, $p <.001$) and social media ($b= 8.48$, $p <.003$) as well as cryptocurrency ($b =15.41$, $p <.001$) were rated as significantly more polarizing. The effect was much larger in the health domain than in the technology domain.

We also asked participants to rate salience as a control factor (Figure 2, right plot). Here, we found that vaccine-related topics were rated higher than coffee (vaccines: $b= 33.06$, $p <.0001$; covid: $b= 27.96$, $p <.0001$). For social media salience was higher than for internet, ($b= 7.98$ $p = 0.067$), while the salience for cryptocurrencies was not statistically different from the one of internet ($b = -4.52$, $p= 0.626$). This should be kept in mind while interpreting the results above.

FIGURE 2

**Discussion**

We confirm our expectations that for most of our queries, academic search engines display scientific literature results in line with the prompted query's directionality, which produces disparities in the evidence that individuals receive, and may amplify confirmation biases. Google Scholar and Semantic Scholar differ in the size of disparities depending on the search query, and we also find differences in disparities between the domains health and technology.

As there is evidence that scientists′ prior beliefs influence their information search and selection (Hergovich et al., 2010; Koehler, 1993; Masnick & Zimmerman, 2009), and confirmation bias has previously been well-established for most individuals seeking information on search engines (Knobloch-Westerwick et al., 2015; Vedejová & Čavojová, 2022), our findings speak to the necessity to conduct more research specifically on the interaction of scientific queries and information searches with academic search engines, an area where research is still lacking: we want to highlight the need for more evidence on how scientific search is conducted on search engines, with a particular focus on how biases are represented and exacerbated (von Hippel & Buck, 2023; White, 2013) and the urgent need for large-scale behavioral data (Khabsa et al., 2016; Rohatgi et al., 2021).

We explored whether we would find patterns with regards to search topics that are polarizing and potentially more salient, as previous literature has established confirmation bias prominent in this regard (Vedejová & Čavojová, 2022; Warner, 2023). Our results neither reject nor confirm that polarization affects result disparities. We find mixed results across topics and search engines: for Google Scholar, search results for the query 'vaccines' were presented with a large disparity, while the query 'covid vaccines' presented more balanced results. For Semantic Scholar, the pattern is reversed to the point in which the searching for 'covid vaccine risks' displays substantially more benefits than searching for benefits. To a small degree, this unexpected result is also observed for coffee. A possible reason is that search engines substantially weight the appearance of the specific words of the search query (in this case, "risks") in the content (Hussan, 2020) disproportionally increasing the rank, even though the search result confirm benefits (potentially without explicitly using the word benefit).

For the technology domain, we found all queries to have similarly large results disparities, with the exception of the query 'social media' in Google Scholar. More research will need to be conducted, both on how topic polarization and salience affect search results in search engines, but also on whether individuals' search queries and behaviors are affected and differ when topic are more controversial. A particular focus should be on the distinction between how scientific novices and scientifically highly educated individuals search for sources and evidence, a field that has not so far received much attention (Hergovich et al., 2010; Krems & Zierer, 1994).

In terms of a limitation of the study design, we would like to discuss the issue of the absence of a pre-defined baseline against which we could measure the skew of search results. In our study, we present disparities between results generated by queries framed in a positively biased (benefits) or negatively biased (risks) manner, as a distance between the two. Indeed, a "skew" measure against a baseline could be established, however, optimally, this would be measured as a deviation from an unbiased baseline. This implies a comprehensive understanding of the true distribution of evidence on a topic, ideally based upon an aggregation of all relevant scientific evidence. The inclusion of a neutral query alone would not necessarily provide a reliable baseline: neutral queries can also produce skewed results if the search engine's algorithms or the underlying indexed content are biased. Despite this challenge, including a baseline query in our study would have had several benefits. Even if the baseline query itself returned skewed results itself, it could still provide a useful reference point to help to contextualize the disparities observed between positively and negatively biased queries, allowing us to assess whether these disparities are particularly pronounced, and whether search engine results tend to converge towards a middle ground when queried in a more neutral manner.

As is, our study does not allow any interpretation regarding how far these results deviate from a neutral query baseline, nor how they compare with the distribution of evidence outside the context of search engines. We argue, however, that the disparities observed between positive and negative queries reflect a maximum range of skewness of either query, if we assume that any neutral query should be between the extremes. If a neutral query lies beyond either, this would suggest an even more problematic issue with the display of search results in academic search engines in general. Future research could tackle the challenge of building semantically well-defined neutral queries. For example, using simply "coffee", or "coffee effects", or even both directionalities, "coffee benefits and risks" or "coffee risks and benefits" are candidates that could be tested. Future audits could also use queries for which some baselines have been established in the scientific literature through for example systematic reviews.

We point to several other limitations. First, we were constrained in our audit to two domains (health and technology) and a limited number (3) of queries per topic, for cost reasons. Wider-ranging generalizations should be made with care, and more research is needed on a wider variety of topics to confirm our findings. Second, we only compare Google Scholar and Semantic Scholar, as two of the most popular academic search engines; our results cannot speak for other scientific search engines or databases such as Scopus or Web of Science. As Google Scholar's results are not reproducible across identical queries (Gusenbauer & Haddaway, 2020), a particular focus of future research could be on how this affects scientific inquiry. Third, for ratings of risks and/or benefits, while full articles might discuss both risks and benefits more broadly, we used the abstract to decide about article valence. We work on the assumption that abstracts contain key takeaways of scientific articles and are the main content observed when screening literature (Macedo-Rouet et al., 2012).

In conclusion, this study underscores the critical need for increased awareness and scrutiny of the biases present within academic search engines. As these platforms play a pivotal role in shaping the information landscape for researchers, educators, and students (Ortega, 2014; Phillips et al., 2002), it is imperative that they provide access to a diverse and balanced range of scientific literature. The results disparity in the evidence base available to users who search with biased queries may have far-reaching consequences for decision-making, policy formulation, and the advancement of knowledge. For scientists and researchers, this research emphasizes the importance of being cognizant of the potential biases inherent in their information-seeking behaviors. Understanding how search engines may inadvertently reinforce pre-existing beliefs is crucial for maintaining objectivity and rigor in the scientific process (von Hippel & Buck, 2023). Researchers should consider adopting strategies to mitigate confirmation bias, such as deliberately using a variety of search query formulations, seeking out diverse sources of information and critically evaluating search results. Search engines should be adapted to aid researchers in this process by providing alerts for skewed queries, which can be identified through recent advancement in natural language processing.

FIGURES

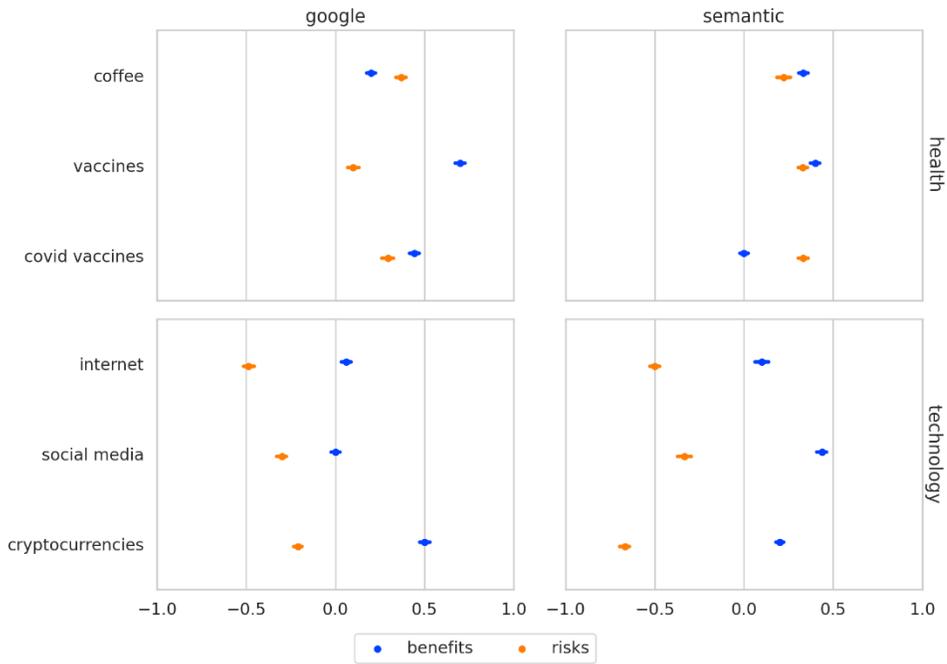

*Figure 1. Disparities in results shown as the distance between queries biased towards benefits (blue dot) and risks (red dot) for each domain (health top, technology use bottom) and for two search engines (Google Scholar left, Semantic Scholar right). X-axis is the dependent variable: Dots further right indicate abstracts disparate towards benefits, dots further left indicate abstracts disparate towards risks.*

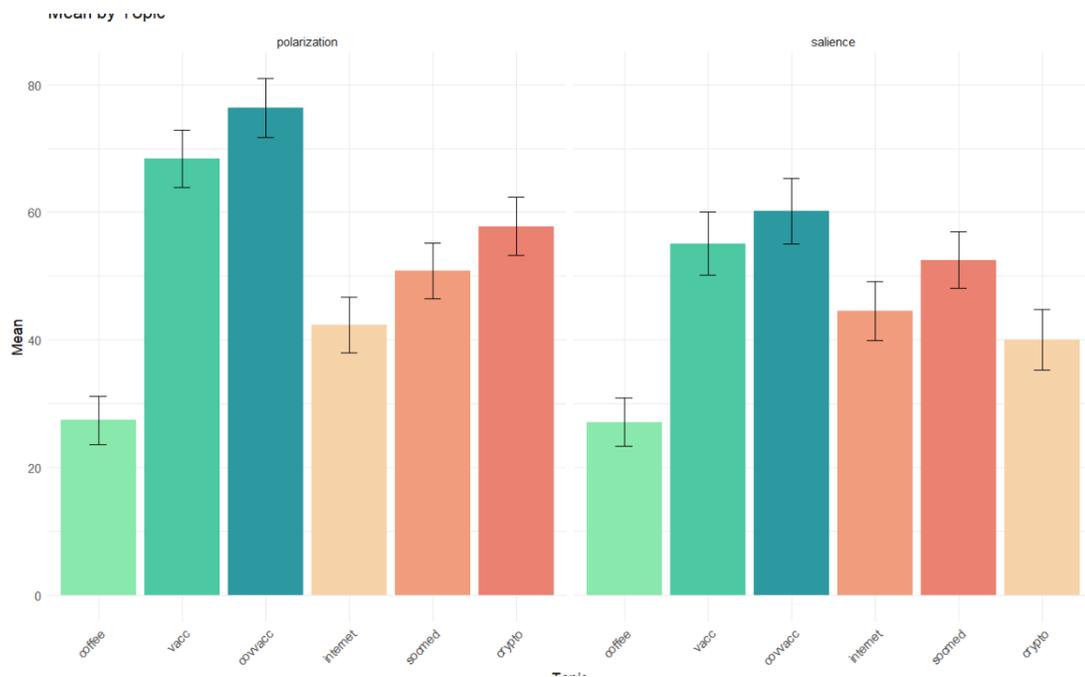

*Figure 2. Ratings of topics for health (teal) and technology use (orange) on polarization and salience. Color shade indicates lowest to highest mean rating within category and rating scale.*

DECLARATIONS

**FUNDING**

The authors declare that they have no known competing financial interests that could have appeared to influence the work reported in this paper.

**CONFLICTS OF INTEREST/COMPETING INTERESTS**

The authors declare that they have no known conflicts of interest or personal relationships that could have appeared to influence the work reported in this paper.

**DATA AVAILABILITY**

Data can be accessed from the repository at: https://zenodo.org/doi/10.5281/zenodo.10636246

**CODE AVAILABILITY**

Code can be accessed from the repository at: https://github.com/robertour/search-scholar

**AUTHORS' CONTRIBUTIONS**

**CK**: Conceptualization, Methodology, Data curation, Formal analysis, Writing- Original draft, Writing – Review & Editing. **RU**: Conceptualization, Methodology, Data curation, Formal analysis, Writing- Original draft, Writing – Review & Editing. **MB**: Data Curation, Writing – Review & Editing. **MM**: Data Curation, Writing – Review & Editing. **MS**: Data Curation.

# Supplementary Materials

# 1. Coverage Tables
## 1.1. Google Scholar

| Google Scholar | day | 1 | | 2 | | 3 | | 4 | | 5 | | 6 | | 7 | | 8 | |
|---|---|---|---|---|---|---|---|---|---|---|---|---|---|---|---|---|---|
| | region | de | oh | de | oh | de | oh | de | oh | de | oh | de | oh | de | oh | de | oh |
| query | browser | | | | | | | | | | | | | | | | |
| coffee consumption health benefits | chrome | 3 | 4 | 3 | 4 | 3 | 4 | 4 | 4 | 3 | 4 | 4 | 4 | 4 | 4 | 4 | 4 |
| | firefox | 3 | 4 | 3 | 4 | 3 | 4 | 2 | 4 | 3 | 4 | 2 | 4 | 2 | 4 | 2 | 4 |
| coffee consumption health risks | chrome | 3 | 4 | 3 | 4 | 3 | 4 | 4 | 4 | 3 | 4 | 4 | 4 | 4 | 4 | 4 | 4 |
| | firefox | 3 | 4 | 3 | 4 | 3 | 4 | 2 | 4 | 3 | 4 | 2 | 4 | 2 | 4 | 2 | 4 |
| covid vaccines health benefits | chrome | 3 | 4 | 3 | 4 | 3 | 4 | 4 | 4 | 3 | 4 | 4 | 4 | 4 | 4 | 4 | 4 |
| | firefox | 3 | 4 | 3 | 4 | 3 | 4 | 2 | 4 | 3 | 4 | 2 | 4 | 2 | 4 | 2 | 4 |
| covid vaccines health risks | chrome | 3 | 4 | 3 | 4 | 3 | 4 | 4 | 4 | 3 | 4 | 4 | 4 | 4 | 4 | 4 | 4 |
| | firefox | 3 | 4 | 3 | 4 | 3 | 4 | 2 | 4 | 3 | 4 | 2 | 4 | 2 | 4 | 2 | 4 |
| cryptocurrency use benefits | chrome | 3 | 4 | 3 | 4 | 3 | 4 | 4 | 4 | 3 | 4 | 4 | 4 | 4 | 4 | 4 | 4 |
| | firefox | 3 | 4 | 3 | 4 | 3 | 4 | 2 | 4 | 3 | 4 | 2 | 4 | 2 | 4 | 2 | 4 |
| cryptocurrency use risks | chrome | 3 | 4 | 3 | 4 | 3 | 4 | 4 | 4 | 3 | 4 | 4 | 4 | 4 | 4 | 4 | 4 |
| | firefox | 3 | 4 | 3 | 4 | 3 | 4 | 2 | 4 | 3 | 4 | 2 | 4 | 2 | 4 | 2 | 4 |
| internet use benefits | chrome | 3 | 4 | 3 | 4 | 3 | 4 | 4 | 4 | 3 | 4 | 4 | 4 | 4 | 4 | 4 | 4 |
| | firefox | 3 | 4 | 3 | 4 | 3 | 4 | 2 | 4 | 3 | 4 | 2 | 4 | 2 | 4 | 2 | 4 |
| internet use risks | chrome | 3 | 4 | 3 | 4 | 3 | 4 | 4 | 4 | 3 | 4 | 4 | 4 | 4 | 4 | 4 | 4 |
| | firefox | 3 | 4 | 3 | 4 | 3 | 4 | 2 | 4 | 3 | 4 | 2 | 4 | 2 | 4 | 2 | 4 |
| social media use benefits | chrome | 3 | 4 | 3 | 4 | 3 | 4 | 4 | 4 | 3 | 4 | 4 | 4 | 4 | 4 | 4 | 4 |
| | firefox | 3 | 4 | 3 | 4 | 3 | 4 | 2 | 4 | 3 | 4 | 2 | 4 | 2 | 4 | 2 | 4 |
| social media use risks | chrome | 3 | 4 | 3 | 4 | 3 | 4 | 4 | 4 | 3 | 4 | 4 | 4 | 4 | 4 | 4 | 4 |
| | firefox | 3 | 4 | 3 | 4 | 3 | 4 | 2 | 4 | 3 | 4 | 2 | 4 | 2 | 4 | 2 | 4 |
| vaccines health benefits | chrome | 3 | 4 | 3 | 4 | 3 | 4 | 4 | 4 | 3 | 4 | 4 | 4 | 4 | 4 | 4 | 4 |
| | firefox | 3 | 4 | 3 | 4 | 3 | 4 | 2 | 4 | 3 | 4 | 2 | 4 | 2 | 4 | 2 | 4 |
| vaccines health risks | chrome | 3 | 4 | 3 | 4 | 3 | 4 | 4 | 4 | 3 | 4 | 4 | 4 | 4 | 4 | 4 | 4 |
| | firefox | 3 | 4 | 3 | 4 | 3 | 4 | 2 | 4 | 3 | 4 | 2 | 4 | 2 | 4 | 2 | 4 |

*Table S1.1. Google Scholar coverage. Each number in the cell corresponds to the number of agents that successfully collected data for each of the conditions given by the day (first row), region (second row), query (first column) and browser (2nd column).*

## 1.2. Semantic Scholar

| Semantic Scholar | day | 1 | | 2 | | 3 | | 4 | | 5 | | 6 | | 7 | | 8 | |
|---|---|---|---|---|---|---|---|---|---|---|---|---|---|---|---|---|---|
| | region | de | oh | de | oh | de | oh | de | oh | de | oh | de | oh | de | oh | de | oh |
| query | browser | | | | | | | | | | | | | | | | |
| coffee consumption health benefits | chrome | 4 | 4 | 2 | 3 | 4 | 4 | 3 | 4 | 4 | 4 | 4 | 4 | 4 | 4 | 4 | 4 |
| | firefox | 4 | 4 | 4 | 2 | 4 | 4 | 4 | 4 | 4 | 4 | 4 | 4 | 4 | 4 | 4 | 4 |
| coffee consumption health risks | chrome | 4 | 4 | 0 | 0 | 4 | 4 | 4 | 4 | 4 | 4 | 4 | 4 | 4 | 4 | 4 | 4 |
| | firefox | 4 | 4 | 0 | 0 | 4 | 4 | 4 | 4 | 4 | 4 | 4 | 4 | 4 | 4 | 4 | 4 |
| covid vaccines health benefits | chrome | 4 | 4 | 4 | 4 | 4 | 4 | 4 | 4 | 4 | 4 | 4 | 4 | 4 | 4 | 4 | 4 |
| | firefox | 4 | 4 | 4 | 4 | 4 | 4 | 4 | 4 | 4 | 4 | 4 | 4 | 4 | 4 | 4 | 4 |
| covid vaccines health risks | chrome | 4 | 4 | 4 | 4 | 4 | 4 | 4 | 4 | 4 | 4 | 4 | 4 | 4 | 4 | 4 | 4 |
| | firefox | 4 | 4 | 4 | 4 | 4 | 4 | 4 | 4 | 4 | 4 | 4 | 4 | 4 | 4 | 4 | 4 |
| cryptocurrency use benefits | chrome | 4 | 4 | 4 | 4 | 4 | 4 | 4 | 4 | 4 | 4 | 4 | 4 | 4 | 4 | 4 | 4 |
| | firefox | 4 | 4 | 4 | 4 | 4 | 4 | 4 | 4 | 4 | 4 | 4 | 4 | 4 | 4 | 4 | 4 |
| cryptocurrency use risks | chrome | 4 | 4 | 4 | 4 | 4 | 4 | 4 | 4 | 4 | 4 | 4 | 4 | 4 | 4 | 4 | 4 |
| | firefox | 4 | 4 | 4 | 4 | 4 | 4 | 4 | 4 | 4 | 4 | 4 | 4 | 4 | 4 | 4 | 4 |
| internet use benefits | chrome | 4 | 4 | 4 | 4 | 4 | 4 | 4 | 4 | 4 | 4 | 4 | 4 | 4 | 4 | 4 | 4 |
| | firefox | 4 | 4 | 4 | 4 | 4 | 4 | 4 | 4 | 4 | 4 | 4 | 4 | 4 | 4 | 4 | 4 |
| internet use risks | chrome | 4 | 4 | 4 | 4 | 4 | 4 | 4 | 4 | 4 | 4 | 4 | 4 | 4 | 4 | 4 | 4 |
| | firefox | 4 | 4 | 4 | 4 | 4 | 4 | 4 | 4 | 4 | 4 | 4 | 4 | 4 | 4 | 4 | 4 |
| social media use benefits | chrome | 4 | 4 | 4 | 4 | 4 | 4 | 4 | 4 | 4 | 4 | 4 | 4 | 4 | 4 | 4 | 4 |
| | firefox | 4 | 4 | 4 | 4 | 4 | 4 | 4 | 4 | 4 | 4 | 4 | 4 | 4 | 4 | 4 | 4 |
| social media use risks | chrome | 4 | 4 | 4 | 4 | 4 | 4 | 4 | 4 | 4 | 4 | 4 | 4 | 4 | 4 | 4 | 4 |
| | firefox | 4 | 4 | 3 | 4 | 4 | 4 | 4 | 4 | 4 | 4 | 4 | 4 | 4 | 4 | 4 | 4 |
| vaccines health benefits | chrome | 4 | 4 | 4 | 4 | 4 | 4 | 4 | 4 | 4 | 4 | 4 | 4 | 4 | 4 | 4 | 4 |
| | firefox | 4 | 4 | 4 | 4 | 4 | 4 | 4 | 4 | 4 | 4 | 4 | 4 | 4 | 4 | 4 | 4 |
| vaccines health risks | chrome | 4 | 4 | 4 | 4 | 4 | 4 | 4 | 4 | 4 | 4 | 4 | 4 | 2 | 3 | 4 | 4 |
| | firefox | 4 | 4 | 4 | 4 | 4 | 4 | 4 | 4 | 4 | 4 | 3 | 4 | 4 | 4 | 4 | 4 |

Table S1.2. Semantic Scholar coverage. Each number in the cell corresponds to the number of agents that successfully collected data for each of the conditions given by the day (first row), region (second row), query (first column) and browser (2nd column).

## 2. Linear model and contrast outputs

### 2.1.     For valence from treatment, topic and engine

```
fit <- lmer(valence ~ trt * topic * engine + region + browser + (day|id) , data = df) print(sum-
mary(fit))
```

```
Scaled residuals:
    Min      1Q  Median      3Q     Max
-2.5669 -0.6216 -0.4012  0.9252  2.7854

Random effects:
 Groups   Name        Variance  Std.Dev.     Corr
 id       (Intercept) 2.010e-09 0.000044836
          day         3.814e-11 0.000006176 -1.00
 topic    (Intercept) 6.282e-03 0.079260557
 Residual             2.702e-01 0.519835004
Number of obs: 27677, groups:  id, 240; topic, 6

Fixed effects:
                                          Estimate  Std. Error          df t value Pr(>|t|)
(Intercept)                              0.2584654   0.0468814   4.4069287   5.513 0.003961 **
trtbenefits                              0.1959285   0.0127934 27663.1242549  15.315  < 2e-16 ***
topictechnology                         -0.5832381   0.0659684   4.3193369  -8.841 0.000632 ***
enginesemantic                           0.0451475   0.0128889 27663.4478365   3.503 0.000461 ***
regionoh                                -0.0112651   0.0062850 27661.1062097  -1.792 0.073083 .
browserfirefox                          -0.0007273   0.0062617 27661.0902633  -0.116 0.907532
trtbenefits:topictechnology              0.3225238   0.0180145 27663.0512081  17.904  < 2e-16 ***
trtbenefits:enginesemantic              -0.2548513   0.0178898 27663.0751120 -14.246  < 2e-16 ***
topictechnology:enginesemantic          -0.2111949   0.0179426 27663.3542686 -11.771  < 2e-16 ***
trtbenefits:topictechnology:enginesemantic 0.4790486 0.0250381 27663.0675991  19.133  < 2e-16 ***
---
Signif. codes:  0 '***' 0.001 '**' 0.01 '*' 0.05 '.' 0.1 ' ' 1

Correlation of Fixed Effects:
            (Intr) trtbnf qry_ty engnsm reginh brwsrf trtb:_ trtbn: qry_t:
trtbenefits -0.139
qry_typtchn -0.704  0.099
enginesmntc -0.139  0.505  0.098
regionoh    -0.074  0.000  0.000  0.034
browserfrfx -0.059  0.000  0.000 -0.020 -0.039
trtbnfts:q_  0.099 -0.710 -0.138 -0.359  0.000  0.000
trtbnfts:ng  0.099 -0.715 -0.071 -0.719  0.001  0.000  0.508
qry_typtch:  0.099 -0.363 -0.138 -0.717  0.000  0.001  0.506  0.516
```

```
    trtbnfts:_: -0.071  0.511  0.099  0.514  0.000  0.000 -0.719 -0.715 -0.716
    optimizer (nloptwrap) convergence code: 0 (OK)
```

## 2.2.       For valence from treatment, Google Scholar specific contrasts and issues

```
fit <- lmer(valence ~ trt * issue + region + browser + (day|id), data = df)) print(summary(fit))

Formula: valence ~ trt * issue + region + browser + (day | id)
   Data: df

REML criterion at convergence: 18025.7

Scaled residuals:
    Min      1Q  Median      3Q     Max
-2.7589 -0.7581 -0.1054  0.6550  2.1306

Random effects:
 Groups   Name        Variance  Std.Dev.  Corr
 id       (Intercept) 0.000e+00 0.000e+00
          day         2.077e-19 4.557e-10 NaN
 Residual             2.250e-01 4.744e-01
Number of obs: 13328, groups:  id, 112

Fixed effects:
                                 Estimate Std. Error         df t value Pr(>|t|)
(Intercept)                     3.837e-01  1.534e-02  1.331e+04  25.017  < 2e-16 ***
trtbenefits                    -1.696e-01  2.005e-02  1.331e+04  -8.463  < 2e-16 ***
issuecovid vaccines            -7.500e-02  2.060e-02  1.331e+04  -3.642 0.000272 ***
issuecryptocurrencies          -5.786e-01  2.005e-02  1.331e+04 -28.862  < 2e-16 ***
issueinternet                  -8.554e-01  2.005e-02  1.331e+04 -42.670  < 2e-16 ***
issuesocial media              -6.696e-01  2.005e-02  1.331e+04 -33.405  < 2e-16 ***
issuevaccines                  -2.696e-01  2.005e-02  1.331e+04 -13.451  < 2e-16 ***
regionoh                       -2.415e-02  8.332e-03  1.331e+04  -2.898 0.003757 **
browserfirefox                 -6.581e-04  8.267e-03  1.331e+04  -0.080 0.936556
trtbenefits:issuecovid vaccines 3.179e-01  2.874e-02  1.331e+04  11.060  < 2e-16 ***
trtbenefits:issuecryptocurrencies 8.786e-01 2.835e-02  1.331e+04  30.991  < 2e-16 ***
trtbenefits:issueinternet       7.161e-01  2.835e-02  1.331e+04  25.259  < 2e-16 ***
trtbenefits:issuesocial media   4.696e-01  2.835e-02  1.331e+04  16.566  < 2e-16 ***
trtbenefits:issuevaccines       7.696e-01  2.835e-02  1.331e+04  27.149  < 2e-16 ***
---
Signif. codes:  0 '***' 0.001 '**' 0.01 '*' 0.05 '.' 0.1 ' ' 1
```

### 2.3. Contrasts for the issues  GS

```
           issue    trt_pairwise estimate     SE  df asymp.LCL asymp.UCL  z.ratio p.value sig
1       vaccines risks - benefits  -0.6000 0.0200 Inf   -0.6529   -0.5471 -29.9313  <0.001 ***
2   social media risks - benefits  -0.3000 0.0200 Inf   -0.3529   -0.2471 -14.9657  <0.001 ***
3       internet risks - benefits  -0.5464 0.0200 Inf   -0.5993   -0.4935 -27.2589  <0.001 ***
4 cryptocurrencies risks - benefits -0.7089 0.0200 Inf   -0.7618   -0.6560 -35.3653  <0.001 ***
5 covid vaccines risks - benefits  -0.1482 0.0206 Inf   -0.2025   -0.0939  -7.1965  <0.001 ***
6         coffee risks - benefits   0.1696 0.0200 Inf    0.1168    0.2225   8.4627  <0.001 ***
```

### 2.4. For valence from treatment, Semantic Scholar specific contrasts and issues

```
Formula: valence ~ trt * issue + region + browser + (day | id)
   Data: df

REML criterion at convergence: 22263.8

Scaled residuals:
    Min     1Q  Median     3Q    Max
-2.3322 -0.6362 -0.3818  1.0728  2.5442

Random effects:
 Groups   Name        Variance  Std.Dev.  Corr
 id       (Intercept) 0.000e+00 0.000e+00
          day         1.612e-15 4.015e-08  NaN
 Residual             2.748e-01 5.242e-01
Number of obs: 14349, groups:  id, 128

Fixed effects:
                                  Estimate Std. Error        df t value Pr(>|t|)
(Intercept)                      2.219e-01  1.763e-02 1.433e+04  12.586  < 2e-16 ***
trtbenefits                      1.111e-01  2.287e-02 1.433e+04   4.859 1.19e-06 ***
issuecovid vaccines              1.111e-01  2.261e-02 1.433e+04   4.915 8.98e-07 ***
issuecryptocurrencies           -8.898e-01  2.261e-02 1.433e+04 -39.357  < 2e-16 ***
issueinternet                   -7.222e-01  2.207e-02 1.433e+04 -32.719  < 2e-16 ***
issuesocial media               -5.556e-01  2.265e-02 1.433e+04 -24.529  < 2e-16 ***
issuevaccines                    1.075e-01  2.278e-02 1.433e+04   4.721 2.37e-06 ***
regionoh                         4.274e-04  8.752e-03 1.433e+04   0.049    0.961
browserfirefox                   1.492e-04  8.752e-03 1.433e+04   0.017    0.986
trtbenefits:issuecovid vaccines -4.444e-01  3.124e-02 1.433e+04 -14.226  < 2e-16 ***
trtbenefits:issuecryptocurrencies 7.564e-01  3.124e-02 1.433e+04  24.213  < 2e-16 ***
trtbenefits:issueinternet        4.889e-01  3.086e-02 1.433e+04  15.844  < 2e-16 ***
trtbenefits:issuesocial media    6.597e-01  3.127e-02 1.433e+04  21.097  < 2e-16 ***
```

```
trtbenefits:issuevaccines         -4.085e-02  3.136e-02  1.433e+04  -1.303    0.193
---
Signif. codes:  0 '***' 0.001 '**' 0.01 '*' 0.05 '.' 0.1 ' ' 1
```

## 2.5.    Contrasts for the issues SS

```
            topic    trt_pairwise estimate     SE  df asymp.LCL asymp.UCL   z.ratio p.value sig
1        vaccines risks - benefits  -0.0703 0.0215 Inf   -0.1269   -0.0136  -3.2726   0.006  **
2    social media risks - benefits  -0.7708 0.0213 Inf   -0.8271   -0.7146 -36.1353  <0.001 ***
3        internet risks - benefits  -0.6000 0.0207 Inf   -0.6547   -0.5453 -28.9575  <0.001 ***
4 cryptocurrencies risks - benefits  -0.8675 0.0213 Inf   -0.9237   -0.8114 -40.7527  <0.001 ***
5  covid vaccines risks - benefits   0.3333 0.0213 Inf    0.2772    0.3895  15.6584  <0.001 ***
6          coffee risks - benefits  -0.1111 0.0229 Inf   -0.1714   -0.0508  -4.8593  <0.001 ***
```